\documentclass[a4paper,fleqn,usenatbib]{mnras}

\usepackage{newtxtext,newtxmath}

\usepackage[T1]{fontenc}
\usepackage{ae,aecompl}

\usepackage{graphicx}	

\title[Quenching in Milky Way analogues]{Quenching of satellite galaxies of Milky Way analogues: reconciling theory and observations}

\author[A. S. Font et al.]{
  Andreea S. Font$^{1}$\thanks{E-mail: A.S.Font@ljmu.ac.uk}, Ian G. McCarthy$^1$,
  Vasily Belokurov$^2$, Shaun T. Brown$^1$, Sam G. Stafford$^1$ \\
$^{1}$Astrophysics Research Institute, Liverpool John Moores University, 146 Brownlow Hill, Liverpool L53RF, UK\\
$^{2}$Institute of Astronomy, University of Cambridge, Madingley Road, Cambridge CB3 0HA, UK
}
\date{Accepted XXX. Received YYY; in original form ZZZ}

\pubyear{2020}

\begin{document}
\label{firstpage}
\pagerange{\pageref{firstpage}--\pageref{lastpage}}
\maketitle

\begin{abstract}
The vast majority of low-mass satellite galaxies around the Milky Way and M31 appear virtually devoid of cool gas and show no signs of recent or ongoing star formation.  Cosmological simulations demonstrate that such quenching is expected and is due to the harsh environmental conditions that satellites face when joining the Local Group (LG).  However, recent observations of Milky Way analogues in the SAGA survey present a very different picture, showing the majority of observed satellites to be actively forming stars, calling into question the realism of current simulations and the typicality of the LG.  Here we use the \texttt{ARTEMIS} suite of high-resolution cosmological hydrodynamical simulations to carry out a careful comparison with observations of dwarf satellites in the LG, SAGA, and the Local Volume (LV) survey.  We show that differences between SAGA and the LG and LV surveys, as well as between SAGA and the \texttt{ARTEMIS} simulations, can be strongly reduced by considering differences in the host mass distributions and (more importantly) observational selection effects, specifically that low-mass satellites which have only recently been accreted are more likely to be star-forming, have a higher optical surface brightness, and are therefore more likely to be included in the SAGA survey.  This picture is confirmed using data from the deeper LV survey, which shows pronounced quenching at low masses, in accordance with the predictions of $\Lambda$CDM-based simulations.
\end{abstract}


\section{Introduction}

It is well established that dwarf galaxies present numerous challenges to, and opportunities for refining, theories of galaxy formation (see \citealt{bullock2017} for a recent review). As highly dark matter-dominated systems, they may be used to place some of the strongest constraints on the nature of dark matter (e.g., \citealt{enzi2021,nadler2021}) and, simultaneously, can be used to study a multitude of physical baryonic processes such as feedback and reionization (e.g., \citealt{efstathiou1992,bullock2000,benson2002,mashchenko2008, pontzen2012,wetzel2016,sawala2016}).  

One such physical mechanism that has been studied for some time, but still remains to be fully elucidated, is the quenching of star formation of dwarf galaxies as they become satellites of more massive hosts.  While dwarf galaxies in the field appear to be ubiquitously star-forming \citep{geha2012}, satellite dwarf galaxies are known to be significantly prone to environmentally-driven quenching. Perhaps the most striking case of this is found in the Local Group (LG) pair of galaxies, the Milky Way (MW) and Andromeda (M31), around which the vast majority of dwarf galaxies are depleted of cool gas and have no significant ongoing (or recent) star formation \citep{grcevich2009,spekkens2014,putman2021}. For example, \citet{putman2021} find that 53 out of 55 satellites within the virial radius of the Milky Way have no detectable HI gas, whereas within the virial radius of M31, 27 out of 30 have no detectable gas. Exceptions to these trends include the spectacular Magellanic Clouds, which do show evidence of ongoing star formation (e.g., \citealt{whitney2008,bolatto2011}) and for which their relatively large gravitational masses likely play an important role in their ability to retain gas and maintain star formation following accretion onto the MW's halo.  

Moving beyond the LG, our knowledge of the properties of satellite galaxies of MW `analogues' (systems with similar stellar mass to the MW) has recently grown considerably thanks to several new dedicated surveys of such systems, both in the Local Volume (LV, $\lesssim 10$~Mpc) \citep{karachentsev2013a,danieli2017,smercina2018,bennet2019,bennet2020,crnojevic2019,carlsten2021a,carlsten2021b,muller2019} and at larger distances ($\approx 20-40$~Mpc, the SAGA survey, \citealt{geha2017,mao2021}).  With regards to quenching, an intriguing result from the SAGA survey is that satellites of MW analogues appear to be generally star-forming \citep{geha2017}, which is in marked contrast to that of the LG.  Using the equivalent width of the H$\alpha$ line as an indicator of current star formation, \citet{mao2021} derived the quenched fractions for the satellites of the SAGA hosts. On average, these fractions are quite low relative to the MW.  For example, only $\approx20\%$ of satellites with stellar masses in the range $10^7 -10^8\, {\rm M}_{\odot}$ appear to be quenched in SAGA, which is approximately three to four times lower than that inferred for the MW and M31 (\citealt{wetzel2015}, see also Fig.~\ref{fig:quenched_frac_LG} below).  Similarly low quenched fractions have recently been independently obtained for the SAGA satellites by \citet{karunakaran2021}, who used $UV$ data from GALEX to search for signs of recent star formation in the identified SAGA satellites.  With reinforced evidence that the dwarf galaxies in these MW analogues are star-forming, the question of why there is such a large difference between the LG and these external MW analogues becomes more pressing.  Is the difference due to an unusual environment in the LG compared with the other sites explored so far?  Or perhaps there are unaccounted for observational differences (i.e., in selection) between the LG and analogues samples?

The process of star formation in dwarf galaxies and, ultimately, the cessation of this process, has been studied theoretically for some time.  Computationally speaking, satellite galaxies are challenging to model, as the relevant processes span a large range of dynamical scales and there are still significant uncertainties in the modelling of baryonic physics, particularly on scales that cannot be directly simulated (i.e., `subgrid' physics).  Note that, at least in principle, capturing the effects of environmental processes in the simulations, such as tidal and ram pressure stripping, is more straightforward, as gravitational and hydrodynamical forces are explicitly and self-consistently evaluated in cosmological hydrodynamical simulations (at least above the resolution limit of the simulations).  Thus, so long as the simulated dwarf galaxies have reasonable properties in the field\footnote{Which may, for example, be obtained through calibration of feedback to reproduce certain observables (e.g., the stellar mass--halo mass relation, star formation rates, etc.; see the discussion in \citealt{schaye2015}).}, cosmological hydrodynamical simulations should accurately follow the gravitational and hydrodynamical evolution of galaxies as they become satellites (e.g., \citealt{bahe2015}) and therefore can make useful predictions for the quenched fraction of satellite dwarf galaxies and its dependence on quantities such as satellite mass, host mass, time of accretion, and so on.

Cosmological simulations have been used extensively to study the quenching of satellites around MW-mass galaxies, with the general aim to determine which physical processes are responsible for the observed high quenched fractions in the LG (e.g., \citealt{wheeler2014,fillingham2015,fillingham2016,simpson2018,garrison-kimmel2019,simons2020,akins2021,karunakaran2021}).  Although the emphasis given to different environmental processes may vary between the studies, a general consensus is that MW-like hosts should be highly efficient at quenching the star formation of satellite dwarf galaxies with stellar masses $\lessapprox 10^{7-8}\, {\rm M}_{\odot}$.  From this perspective, and taking into consideration that this picture appears to be confirmed observationally in the LG, the low quenched fractions derived from SAGA are puzzling. Taken at face value, the result implies that the various environmental processes at play must, for some reason, be inefficient in the SAGA hosts (or overly efficient in the LG, if one disregards the predictions of simulations).  Why this should be the case is presently unclear.

The above conclusions are predicated on the assumption that comparisons between theoretical predictions and observations take into account all of the relevant observational selection criteria and that, as far as is possible, the comparisons have been made in a like-with-like manner.  In this vein, in \citet{font2021} we have recently made comparisons between the predictions of the new \texttt{ARTEMIS} suite of cosmological hydrodynamical simulations \citep{font2020} with the observed luminosity and radial distributions functions of satellites in the LG, LV and SAGA samples of MW analogues.  An important aspect of the comparisons in that study was the application of observational selection criteria to the simulations before making those comparisons.  In particular, it was shown that the observed luminosity and radial distribution functions could be reproduced for the LG and MW analogues in the LV survey when applying the observed radial and magnitude selection thresholds from these data sets to \texttt{ARTEMIS}.  However, applying the stated SAGA radial and magnitude limits to \texttt{ARTEMIS} resulted in significantly more satellites in \texttt{ARTEMIS} with respect to SAGA.  Comparing SAGA to the deeper LV survey, \citet{font2021} (see also \citealt{carlsten2021a}) concluded that the SAGA sample has significantly fewer satellites compared to LV in the magnitude range $M_V \approx$ $-12$ to $-14$, where $-12$ is the approximate magnitude limit of SAGA (see figure~1 of \citealt{font2021}).  Specifically, the LV has a relatively large number of satellites in this magnitude range with surface brightnesses lower than $\approx 25$ mag ~arcsec$^{-2}$, whereas such systems are completely absent in SAGA, suggesting that such a surface brightness limit is inherent in the SAGA selection function and/or the spectroscopic follow-up programme.  Applying a surface brightness limit of this order to \texttt{ARTEMIS} yields luminosity and radial distribution functions in excellent agreement with SAGA and provides a simple framework for understanding differences between the LG, LV and SAGA surveys.

An interesting finding of \citet{font2021} is that systems with lower than typical surface brightness (at fixed magnitude) tended to be more centrally concentrated in the \texttt{ARTEMIS} simulations, suggesting an environmental origin to the scatter in surface brightness at fixed luminosity.  If this interpretation is correct, it raises the possibility that the identification/selection of satellites in the observations is itself potentially dependent on the past environmental history of the satellites and may therefore have important implications for comparisons between the predicted and observed quenched fractions of satellites around MW-mass hosts.  

In the present study we make use of the \texttt{ARTEMIS} suite of cosmological hydrodynamical simulations of MW-mass systems to analyse how observational selection effects affect the observed quenched fractions of satellite galaxies in MW analogues.  We compare our simulations with various samples, including those in the LG and from the LV and SAGA surveys.  We demonstrate that, indeed, accounting for observational selection effects is crucially important when making comparisons between the observed and simulated quenched fractions of satellites of MW-mass hosts and that, when this is carefully done, there is no significant tension between the predictions of $\Lambda$CDM-based simulations and current observations of MW analogues, contrary to recent claims.

The present paper is structured as follows.  In Section \ref{sec:sims} we introduce the \texttt{ARTEMIS} simulations and discuss how they are processed.  In Section \ref{sec:obs} we compare the predicted quenched fractions to those derived from various surveys, including the LG (Section \ref{sec:LG}), the SAGA survey (Section \ref{sec:SAGA}) and the LV survey (Section \ref{sec:LV}).  We summarise and discuss our findings in Section \ref{sec:concl}.

\section{Simulations}
\label{sec:sims}

The \texttt{ARTEMIS} simulations were introduced and described in detail in \citet{font2020} (see also \citealt{font2021}), therefore we summarize only the main characteristics and the details pertinent to the present study here. 

The suite comprises $45$ MW-mass galaxies selected to have total masses between $8\times10^{11} < {\rm M}_{200}/{\rm M}_\odot < 2\times10^{12}$, where ${\rm M}_{200}$ is the mass enclosing a mean density of 200 times the critical density at $z=0$.
These systems were selected from a 25 Mpc/$h$ periodic box simulated with collisionless physics and were then re-simulated at higher resolution using the zoom in technique, with full hydrodynamics and a model for galaxy formation (see below).  The simulations are based in a spatially-flat $\Lambda$CDM cosmology using the WMAP-9yr maximum-likelihood parameter values from \citet{hinshaw2013}: $\Omega_m=0.2793$, $\Omega_b=0.0463$, $h=0.70$, $\sigma_8=0.8211$, and $n_s=0.972$. The particle masses are $\simeq 2.2 \times 10^4 \, {\rm M}_{\odot}/h$ for baryons (initially) and $1.17 \times 10^5 \, {\rm M}_{\odot}/h$ for dark matter, while force resolution (i.e., the Plummer-equivalent softening) is $125$ pc/$h$.

The hydrodynamical zoom simulations were run using a version of the Gadget-3 TreePM SPH code \citep{springel2005} developed for the \texttt{EAGLE} project \citep{crain2015,schaye2015}.  The code includes subgrid prescriptions for metal-dependent radiative cooling in the presence of a photo-ionizing UV background, star formation, stellar evolution and chemical enrichment, black hole growth, and feedback associated with star formation and active galactic nuclei (see \citealt{schaye2015} and references therein).  For \texttt{ARTEMIS}, these prescriptions are the same as in the \texttt{EAGLE} simulations, the only exception being that the parameter values associated with the stellar feedback model were adjusted to achieve an improved match to the amplitude of the stellar mass--halo mass relation at the MW halo mass scale (see \citealt{font2020} for details). 

Although the simulations reproduce the amplitude of the stellar mass--halo mass relation by construction, they have not been explicitly tuned to match other galaxy properties.  In spite of this, the simulations do successfully reproduce a broad range of global and structural properties of MW-mass hosts and of their dwarf satellites.  For example, they reproduce the observed sizes and star formation rates of MW-mass galaxies, their magnitudes in different bands, and the spatial distributions of metallicity and mass of stellar haloes \citep{font2020}. They also reproduce the luminosity functions and radial distributions of satellite dwarf galaxies of MW analogues in SAGA and in the LV once important observational selection effects are factored in \citep{font2021}.  These selection effects will also be important for the present study, as we discuss below. 

In order to implement observational selection criteria for the satellites (e.g., magnitude and surface brightness limits) and to facilitate comparisons to the quenched fractions of LV satellites (see Section~\ref{sec:LV}), for which we use the $g-i$ colour as an indicator of quenching\footnote{Spectroscopic-based estimates of the SFR are not yet available for most LV satellites.}, we compute magnitudes and colours of star particles in post-processing using simple stellar population (SSP) models constructed with the \texttt{PARSEC v1.2S}+\texttt{COLIBRI PR16} isochrones\footnote{\url{http://stev.oapd.inaf.it/cgi-bin/cmd}} \citep{bressan2012,marigo2017} and adopting the \citet{chabrier2003} stellar initial mass function (IMF) used in the simulations.  Note that \texttt{COLIBRI} accounts for the thermally-pulsating (TP) AGB phase, from the first TP phase to the complete loss of the envelope.  In order to derive luminosities for satellites in a given band, we simply sum the luminosities of the individual star particles that are gravitationally bound to each satellite, as determined by the \texttt{SUBFIND} algorithm \citep{dolag2009}.  Surface brightnesses, $\mu_e$, represent the \textit{mean} surface brightness within the (projected) effective radius, computed by summing the luminosities of all particles within this radius, dividing through by the enclosed area, and converting the result to standard magnitude units (i.e., mag arcsec$^{-2}$). 

The computed host magnitudes are in very good agreement with those of observed MW analogues in SAGA and LV (see Sections \ref{sec:SAGA} and \ref{sec:LV} and also \citealt{font2021}). The stellar masses and magnitudes of simulated satellites are also realistic (see \citealt{font2021}). In the current study, we use only satellites with $>10$ star particles, which corresponds to stellar masses $\gtrsim 10^{5.5}\, {\rm M}_{\odot}$, or approximately $M_V<-8$.

Lastly, we also constructed the merger trees of the $45$ MW-mass hosts, using the same methods described in \citet{mcalpine2016} for the \texttt{EAGLE} simulations. In this paper, we make use of these merger trees to follow the satellites back in time to determine their redshifts of accretion, which we define as the earliest redshift where a present-day satellite has joined the friends-of-friends group of the main progenitor.
 
\section{Comparison of observed and simulated quenched fractions}
\label{sec:obs}

In this section we compare the star forming properties of satellites in the \texttt{ARTEMIS} simulations with those in the LG (Section \ref{sec:LG}), the SAGA survey (Section \ref{sec:SAGA}), and the LV survey (Section \ref{sec:LV}).

For the comparisons that use the star formation rate directly to assess the quenched fraction (LG and SAGA), we define a galaxy to be quenched if it is not presently forming stars, i.e., if its instantaneous specific star formation rate sSFR$_{\rm inst}$ is zero.  In Appendix~\ref{sec:appendixB} (Fig.~\ref{fig:quenched_frac_sfr}) we explore other definitions, for example by raising slightly the threshold of sSFR$_{\rm inst}$ for quenched galaxies, or by requiring galaxies not to have formed stars over some time window, i.e., we use the time-averaged quantity sSFR$_{\Delta t}$. However, we have verified that these changes do not affect the quenched fractions significantly and we therefore adopt the sSFR$_{\rm inst} =0$ condition by default.

\subsection{Local Group (LG)}
\label{sec:LG}

For comparison to satellites in the LG, we adopt a simple spatial cut to mimic the typical observational selections. Specifically, we include all dwarf galaxies within $300$~kpc of the centre of the host (e.g., \citealt{mcconnachie2012,wetzel2015}). This choice is motivated by the fact that $300$~kpc corresponds to approximately the virial radius of a MW-mass galaxy\footnote{Alternatively, adopting an overdensity of 200 with respect to the critical density yields ${\rm R}_{200} \approx 220$~kpc for a halo with ${\rm M}_{200} \approx 1.4 \times 10^{12} \, {\rm M}_{\odot}$.}. Although the LG observations include ultra-faint dwarfs, we compare only with the `classical' dwarfs (i.e., stellar masses $\gtrsim 10^{5.5} \, {\rm M}_{\odot}$, or $M_V \la -8$), which is the regime resolved in the simulations. In principle, the LG observations are also surface brightness limited, but this limit ($\mu_{0,V} \approx 29-30$ mag arcsec$^{-2}$; see, e.g., \citealt{drlica-wagner2020}) does not affect the regime of dwarfs studied here.  

\begin{figure}
\includegraphics[width=\columnwidth]{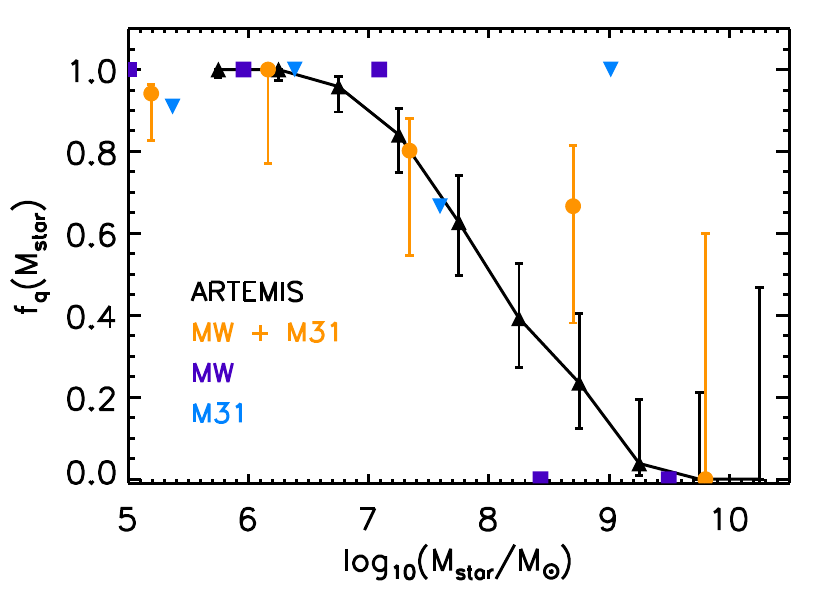}\\
\caption{Mean quenched fraction vs.~stellar mass for satellites with distances $< 300$~kpc in the simulated \texttt{ARTEMIS} MW analogues (black triangles; full curve). The error bars correspond to the 2-sigma confidence interval assuming a binomial beta distribution. For comparison, we show the measured quenched fractions for the combined LG (MW + M31) hosts (orange circles) and also separately for the MW (purple squares) and M31 (blue triangles) satellites using data from \citet{wetzel2015}. The simulations reproduce the average LG trend relatively well.}
\label{fig:quenched_frac_LG}
\end{figure}

Fig.~\ref{fig:quenched_frac_LG} presents the quenched fraction in bins of stellar mass for all satellites with stellar masses $> 10^{5.5}\, {\rm M}_{\odot}$ in the simulations, shown as an average over all simulated MW-mass hosts (black curve and triangles).  The error bars are computed assuming a binomial beta distribution (e.g., \citealt{cameron2011}) and represent the 2-sigma (95\%) confidence interval on the stack, rather than the typical uncertainty for a given host.
For comparison, we show the measured quenched fractions of satellites in the MW and M31 from \citet{wetzel2015}. The latter are shown both combined (MW+M31) and separately for each host.  For the combined MW+M31 result, the error bars are computed by \citet{wetzel2015} also assuming a binomial beta distribution and represent the 1-sigma confidence interval.  The mean quenched fraction in the simulations and observations agree fairly well with each other, even though no aspect of the properties of satellites was calibrated to prior to running the simulations (the feedback was calibrated on the host stellar mass only).  In both cases, low-mass satellites (M$_{\rm star} \lesssim 10^{7}\, {\rm M}_{\odot}$), are almost completely quenched, while the majority of satellites with  M$_{\rm star} \gtrsim 10^{8}\, {\rm M}_{\odot}$ are currently star-forming. We find, however, that there is non-negligible host-to-host scatter in the simulations, of similar magnitude as seen in the individual LG hosts, MW and M31.  

Note that for the MW, the two largest stellar mass bins correspond to the SMC and LMC, which are both star forming.  Why the most massive satellites in M31 appear quenched whereas the SMC/LMC in the MW are not is unclear, but it may be related to differences in masses of the two hosts, with M31 being a factor of two more massive than the MW and therefore having a stronger environmental effect on its satellites.  Alternatively, if the SMC and LMC have only recently been accreted, there may have been insufficient time for environmental processes to quench these satellites. Note also that the compilation data of \citet{wetzel2015} does not include M33. Including this gas-rich, star-forming spiral \citep{kam2017} should lower the quenched fraction somewhat for high mass satellites in M31.

The overall agreement between the simulations and the LG in Fig.~\ref{fig:quenched_frac_LG} is reassuring, as it is in line with results obtained from other recent cosmological hydrodynamical simulations (e.g. \citealt{akins2021,karunakaran2021}). Clearly, the LG environment is conducive to environmental quenching of low-mass satellite galaxies and our simulations, which include a large number of MW-mass hosts, predict that this is the norm, rather than an exception. 

The relatively large samples of observed MW analogues from beyond the LG now allow us to test this prediction. We first compare the simulated quenched fractions with those inferred for satellites of MW analogues in SAGA.

\begin{figure*}
    \centering
    \includegraphics[width=\columnwidth]{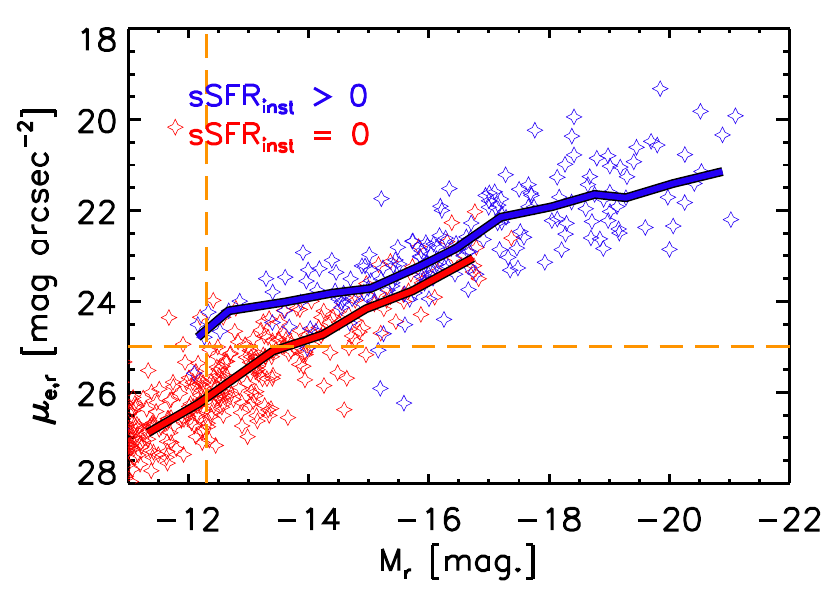}
    \includegraphics[width=\columnwidth]{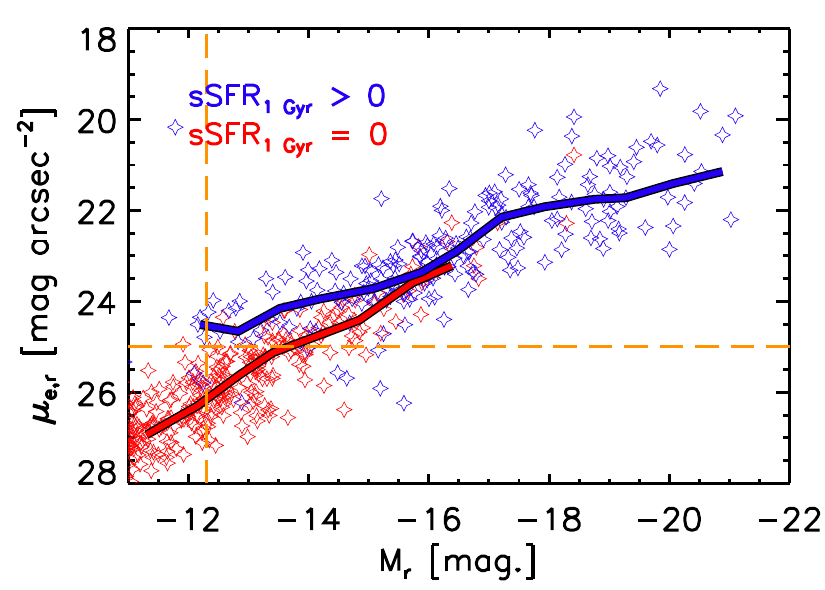}\\
    \includegraphics[width=\columnwidth]{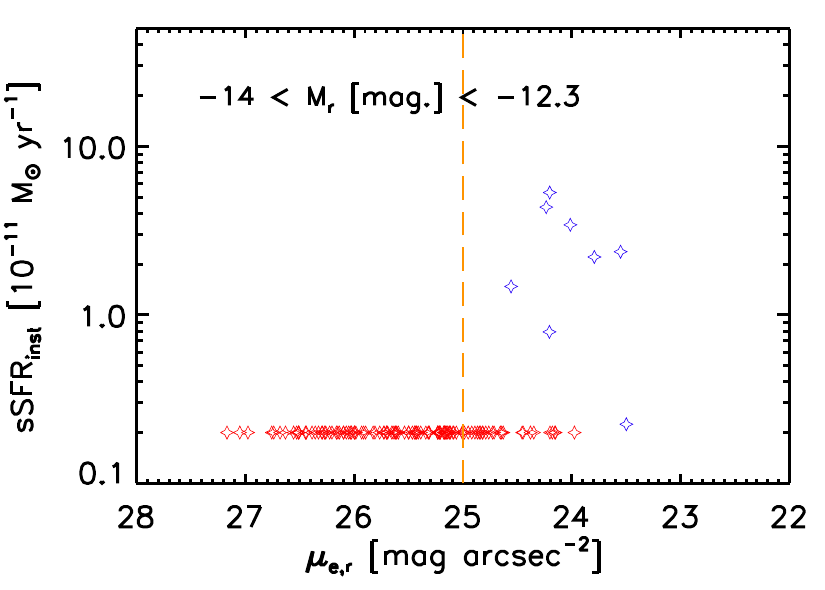}
    \includegraphics[width=\columnwidth]{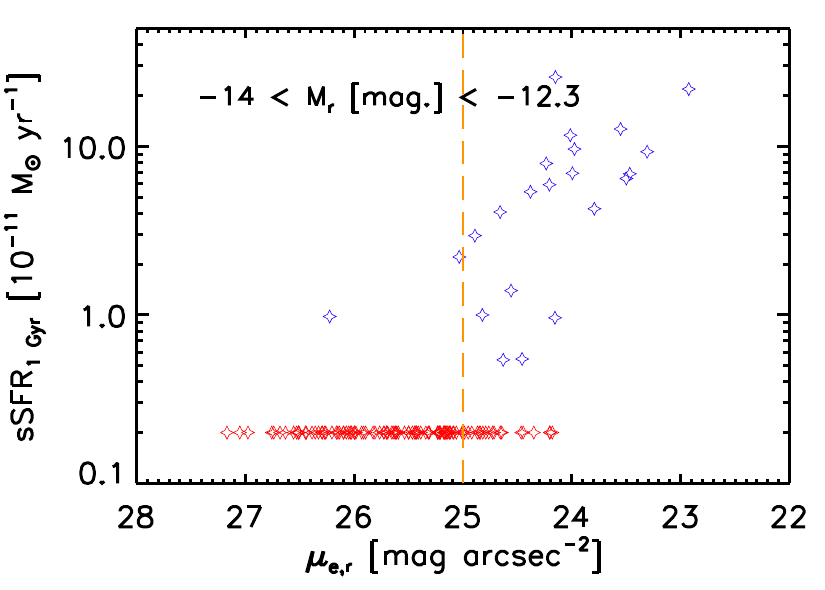}\\    
    \caption{{\it Top:} Magnitude--surface brightness relation of satellites in the \texttt{ARTEMIS} simulations, colour-coded according to star forming status (star-forming or quenched) using two different definitions of the star formation rate: the default instantaneous sSFR ({\it left}) or the average sSFR over the past 1~Gyr ({\it right}).  The solid curves represent the median trends for the quenched and star-forming populations.  The dashed orange lines correspond to the nominal magnitude limit of SAGA ($M_r = -12.3$) and its apparent surface brightness limit of $\mu_{e,r} = 25$ mag arcsec$^{-2}$ based on comparison to LV (see text and \citealt{font2021}).  Relatively bright satellites (e.g., $M_r < -14$) are more likely to be star-forming, whereas relatively faint satellites (e.g., $M_r > -14$) are much more likely to be quenched.  At magnitudes of $-14$ to $-12$, star-forming satellites tend to have higher surface brightnesses than quenched satellites, therefore making them potentially much easier to detect observationally.  {\it Bottom:} The relation between sSFR and surface brightness in the magnitude range $-14$ to $-12.3$ (i.e., the magnitude range where the surface brightness limit comes into play for SAGA), for the two definitions of sSFR.  Note that systems with sSFR=0 have been arbitrarily shifted to sSFR $= 2\times10^{-12}$ M$_\odot$ yr$^{-1}$ in order to display them on a logarithmic axis.  Over this magnitude range, satellites with surface brightness brighter than about 25 mag arcsec$^{-2}$ are likely to be star forming with a surface brightness that correlates with the sSFR (especially the time-averaged definition).} 
    \label{fig:mu_SFR_mag}
\end{figure*}

\subsection{SAGA}
\label{sec:SAGA}

The current SAGA sample (``stage 2'') contains $36$ MW analogues, of which we use the $34$ systems which have identified satellites around them. The mean host K-band magnitude of the entire sample is $M_K=-23.72$ \citep{geha2017,mao2021}. The size of this observational sample is comparable to that of \texttt{ARTEMIS}, which enables a statistical comparison. The simulated hosts have a similar mean K-band luminosity (corresponding to $M_K = -23.84$) but a somewhat different distribution of luminosities from SAGA.  Note that the K-band luminosity is considered as a proxy for stellar mass and we account for the impact of (slightly) different $M_K$ distributions in the simulations and SAGA below.

In SAGA, satellite galaxies are selected with the following criteria: a projected distance of $<300$~kpc from the centre of their hosts, a line of sight velocity of $|{\rm v}_{\rm l. o. s.}|< 250$~km s$^{-1}$, and a magnitude completion limit of $M_r < -12.3$ \citep{geha2017}. However, as described in \citet{font2021}, the SAGA survey is also likely to have a relevant surface brightness limit of $\mu_{e,r} \approx 25 \, {\rm mag} \, {\rm arcsec}^{-2}$ (see figure~1 in that study).  This is a rough estimate of the surface brightness limit based on a comparison with the deeper LV survey, which shows significantly more satellites than SAGA in the magnitude range $-14$ to $-12$, many of which have surface brightnesses fainter than the level mentioned above (noting here that SAGA identified virtually no satellites below this limit).  By including this surface brightness limit when selecting satellites from the \texttt{ARTEMIS} simulations, we demonstrated that it is possible to match a wide range of dwarf galaxy properties in SAGA, namely the abundance of satellites and its dependence on host magnitude, as well as the luminosity functions and radial distributions of ($M_r<-12$) satellites. 

There are two possible sources of incompleteness which could affect the observed quenched fractions: 1) genuine satellites were undetected in the initial imaging data (e.g., due to low surface brightness) and consequently not followed up to assess their star-forming status; and/or 2) detected satellites were not followed up for spectroscopy, or have inconclusive redshifts from the follow-up. The second possibility has been carefully considered via detailed modelling in \citet{mao2021} (see also the red error bars in Fig.~\ref{fig:quenched_frac_SAGA} below) and, while partially alleviating the tension between SAGA and theory, this source of incompleteness clearly cannot reconcile the tension on its own.  We therefore speculate that if incompleteness issues are the culprit, it is likely to require a significant contribution from the first source of incompleteness (non-detection in the initial imaging).  As we discuss below, in the simulations the surface brightness of low-mass satellites is sensitive to environmental processes and may provide an explanation for why quenched satellites in particular are difficult to detect observationally.

With regards to the possibility of undetected low surface brightness satellites, \citet{mao2021} have compared SAGA with the `low surface brightness galaxy' sample independently produced by \citet{tanoglidis2021} using data from the Dark Energy Survey (DES).  For all SAGA systems that fall within the DES footprint, \citet{mao2021} compared the two catalogs, finding that all low surface brightness galaxies in the \citet{tanoglidis2021} catalog were also present in the SAGA catalog and with consistent properties.  We note, however, that \citet{kado-fong2021} estimate the ($g$-band) mean surface brightness completeness limit of the DES data in \citet{tanoglidis2021} to be $\mu_{e,g} \approx 25.75\, {\rm mag} \, {\rm arcsec}^{-2}$ (see section 2.2 and fig.~2 of \citealt{kado-fong2021}) on the basis of comparisons to considerably deeper Hyper Suprime-Cam (HSC) imaging.  Adopting a typical satellite colour of $g-r \approx 0.5$, this corresponds to $\mu_{e,r} \approx 25.25\, {\rm mag} \, {\rm arcsec}^{-2}$ which is close to the (by eye) limit $\mu_{e,r} \approx 25\, {\rm mag} \, {\rm arcsec}^{-2}$ from \citet{font2021}.  We therefore argue that consistency between the SAGA and \citet{tanoglidis2021} photometric catalogs does not preclude the possibility that both catalogs may be missing lower surface brightness satellites.  On the contrary, the comparisons of SAGA with LV in \citet{font2021} and the DES and HSC catalogs in \citealt{kado-fong2021} suggest that this is the case.

In the following, we explore the impact of an additional surface brightness limit as part of the SAGA selection criteria and show how this may affect the deduced quenched fractions of MW analogues in SAGA.  

We begin by examining the top panels of Fig.~\ref{fig:mu_SFR_mag}, which show the magnitude--surface brightness relations of satellites in the \texttt{ARTEMIS} simulations, colour-coded according to star-forming status (i.e., active or quenched).  Here we use two different estimates of the sSFR, which is our indicator of star-forming status.  In particular, we use the default instantaneous value, sSFR$_{\rm inst}$, of the star-forming gas and, for comparison, the average over the past 1 Gyr, sSFR$_{\rm 1 Gyr}$.  Note that to determine the time-averaged sSFR, we simply sum the (initial) masses of all star particles which actually formed within the past 1 Gyr and divide through by that timescale and the current stellar mass of the satellite.

Regardless of which estimate of the sSFR is used, we find that relatively bright satellites (e.g., $M_r \la -14$) are more likely to be star-forming, whereas relatively faint satellites (e.g., $M_r \ga -14$) are much more likely to be quenched.  At magnitudes of $-14$ to $-12$ (i.e., near the SAGA selection threshold), star-forming satellites tend to have higher surface brightnesses than quenched satellites, making them potentially easier to detect observationally.  This is particularly the case if the satellite has had any recent star formation, e.g., within the past 1 Gyr. Notably, this $\sim 1.5$~mag difference in surface brightness
between star-forming and quenched satellites is about the same level as found observationally between late-type and early-type dwarfs in the LV survey (see figure 2 of \citealt{carlsten2021b}).

In the bottom panels of Fig.~\ref{fig:mu_SFR_mag}, we show the relations between sSFR and surface brightness in the magnitude range $-14$ to $-12.3$, again for the two definitions of sSFR.  Over this magnitude range, satellites with surface brightness brighter than about 25 mag arcsec$^{-2}$ are more likely to be star-forming and those that are have surface brightnesses that correlates with the sSFR. We find that 22\% and 15\% of satellites are classified as star-forming above this surface brightness limit when using the time-averaged and instantaneous sSFR definitions, respectively.

Based on the above, the observational selection function of satellites may be strongly affected by star-forming status.  Certainly this is the case in the \texttt{ARTEMIS} simulations, and it underscores the importance of making like-with-like comparisons between the simulations and observations.  We will explore the physical origin of this selection effect below (see Fig.~\ref{fig:mu_zacc}), but first we examine its impact on the observed satellite quenched fractions of MW analogues.

In order to understand how each of the observational selection criteria affects the computed quenched fractions of satellites in the simulations, we analyse them separately, as follows:

\begin{itemize}
    \item {\it `LG selection':} satellites selected within a (3D) distance of $300$~kpc from the centre of their host. We include this as our reference case.  
    \item {\it `SAGA selection':} satellites with projected distances  $<300$~kpc and  line-of-sight velocities, $v_{\rm l. o. s.}\le 250$~km/s and magnitudes $M_r<-12.3$.  Here we project along the z-axis of the simulation box. Note that the projected separation + $v_{\rm l. o. s.}$ criteria is likely to include some fraction of interlopers. As field galaxies are generally star-forming, this is expected to slightly lower the quenched fractions.
    \item {\it `SAGA,$\mu$ selection':} in addition to the stated SAGA criteria, we impose a mean effective $r$-band surface brightness limit of $\mu_{e,r} <25 $ mag arcsec$^{-2}$, as discussed above. 
\end{itemize}

\begin{figure}
\includegraphics[width=\columnwidth]{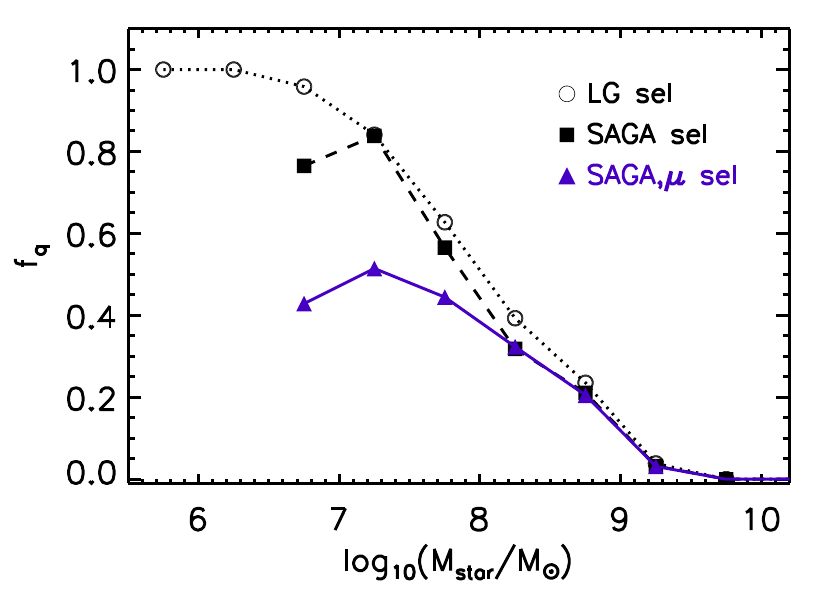}\\
\includegraphics[width=\columnwidth]{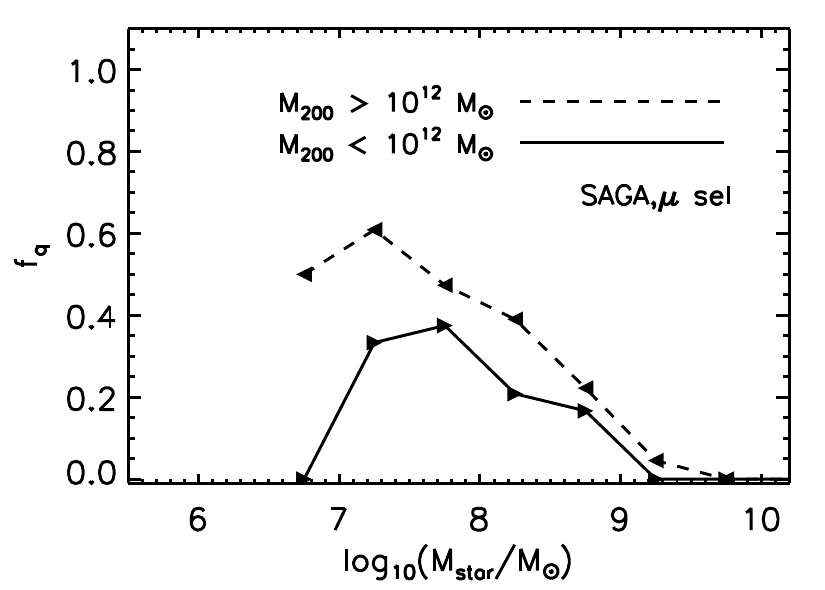}
\caption{{\it Top:} Mean quenched fractions in the simulations after applying different satellite selection criteria: the `LG selection', the `SAGA selection' (i.e., nominal SAGA selection function), and  the `SAGA,$\mu$ selection' (nominal + surface brightness cut). See text for details on these selections. The magnitude cut in the SAGA selection mainly reduces the range in stellar masses compared with the LG selection, while the additional surface brightness cut in SAGA,$\mu$ has the strongest impact on the quenched fraction trend with stellar mass. {\it Bottom:} The mean quenched fraction of satellites with the SAGA,$\mu$ selection, shown separately for two ranges of host masses, above and below $10^{12} \, {\rm M}_{\odot}$.  More massive hosts have higher fractions of quenched satellites of a given stellar mass.}
\label{fig:quenched_frac_var}
\end{figure}

In Fig.~\ref{fig:quenched_frac_var} we show the mean quenched fractions in the simulations, applying the different selection criteria described above.  Comparing first the LG and nominal SAGA selections, we conclude that: i) the magnitude cut for the SAGA selection mainly limits the range of stellar masses that are sampled, specifically increasing the lower bound from $\approx 10^6$ to $\approx 10^7 \, {\rm M}_{\odot}$; and ii) the use of projected radii with a line-of-sight velocity cut results in a slight decrease in the quenched fraction relative to the LG selection at high stellar masses ($\sim 10^8 {\rm M}_{\odot}$), which is due to the inclusion of a small number of star-forming interlopers.  Overall, though, the nominal SAGA selection does not result in a significantly different quenched fraction trend with stellar mass compared to the LG selection.  By contrast, including the surface brightness cut (the SAGA,$\mu$ selection) has a substantial effect on the quenched fractions: at the low-mass end, ${\rm M}_{\rm star} \approx 10^{7}\, {\rm M}_{\odot}$, the quenched fractions decrease from $\approx 85\%$ to $\la 50\%$; compare the SAGA and SAGA, $\mu$ selections.  Therefore, a limiting surface brightness has an important effect on the (de)selection of quenched galaxies, and is therefore able to lower the quenched fractions significantly. 

Aside from the satellite selection criteria, the host mass is also expected to be a relevant factor in the resulting quenched fractions, as dwarf galaxies of a given mass are more likely to be quenched if they reside in a more massive host, since processes such as ram pressure stripping and tidal stripping should be more efficient at higher host masses.  We examine the host mass dependence in the bottom panel of Fig.~\ref{fig:quenched_frac_var}, which shows the mean quenched fractions of satellites selected with the SAGA,$\mu$ selection, separated by the mass of their host. This clearly demonstrates that hosts with total halo masses $>10^{12}\, {\rm M}_{\odot}$ have a higher fraction of quenched satellites (at fixed stellar mass) than those below this host mass threshold. As observational samples of MW analogues also include a variety of host masses, one should expect a mix of quenched fractions.  Thus, a more consistent comparison can be made if both observational and simulated samples are scaled to a common host mass, which we aim to achieve by using the K-band luminosity as a proxy for host mass. Note, however,that because there is significant intrinsic scatter in the host luminosity--total mass relation, splitting the sample by luminosity (as opposed to halo mass) will generally not result in as large an effect as seen in the bottom panel of Fig.~\ref{fig:quenched_frac_var}.  Nevertheless, scaling to a common host luminosity is clearly better than taking no account of the differences in the host samples.  In the future, it may be possible to use dynamics-based indicators of the host masses, as opposed to the luminosity.

We investigate first the un-scaled quenched fractions, now comparing \texttt{ARTEMIS} with the SAGA survey. The top panel of Fig.~\ref{fig:quenched_frac_SAGA} shows the mean quenched fractions in the simulations, using our adopted SAGA,${\mu}$ selection, compared with that derived from the SAGA data by \citet{mao2021}. The red error bars, also from \citet{mao2021}, show the maximum quenched fraction were all the satellites not followed up spectroscopically to be quenched.  The orange error bars correspond to the 1-sigma shot noise uncertainties derived by \citet{mao2021}, whereas the black error bars on the simulation data points correspond to 2-sigma uncertainties.  In general, there is a reasonable agreement with the observations, with \texttt{ARTEMIS} showing only slightly elevated quenched fractions with respect to SAGA.  This should be contrasted with the findings of \citet{karunakaran2021}, who find much larger tensions between SAGA and the Auriga and \texttt{APOSTLE} simulations. Note that the trend here for \texttt{ARTEMIS} differs slightly from that shown in Fig.~\ref{fig:quenched_frac_var} due to differences in the stellar mass binning, as we have attempted to match the binning strategy of \citet{mao2021} for the comparison in Fig.~\ref{fig:quenched_frac_SAGA}.

\begin{figure}
\includegraphics[width=\columnwidth]{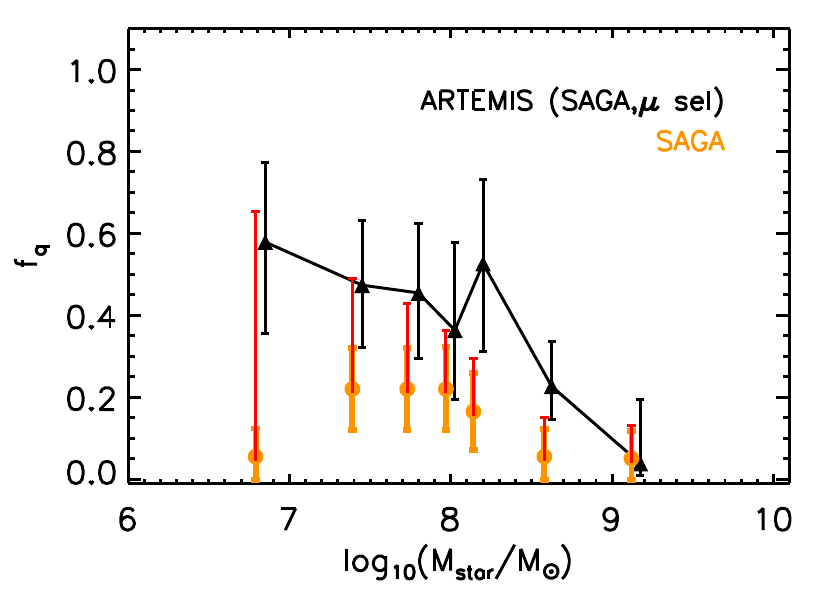}\\
\includegraphics[width=\columnwidth]{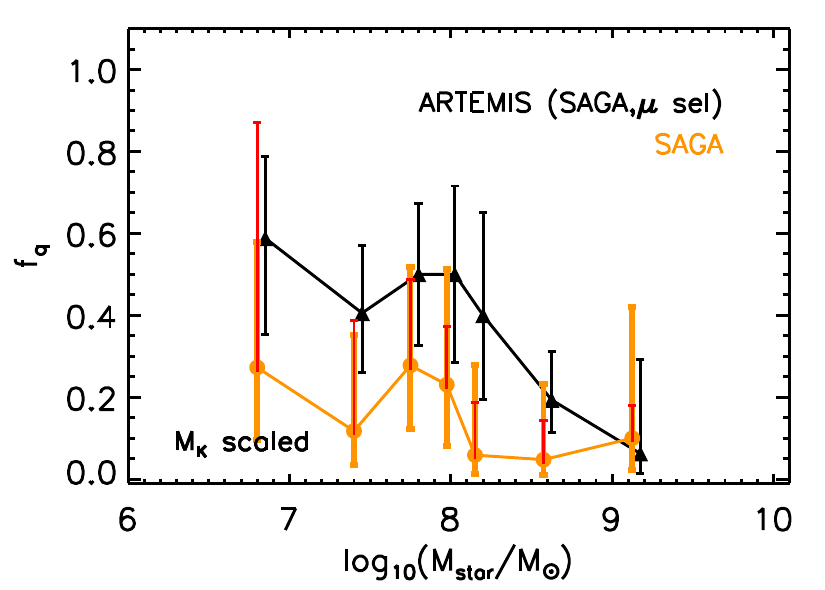}
\caption{{\it Top:} Mean quenched fractions in the simulations using the SAGA,$\mu$ selection (black triangles and black curve).   The quenched fractions from SAGA \citep{mao2021} are shown for comparison, with both statistical shot noise error bars (orange) and maximal systematic errors accounting for satellites not followed up spectroscopically (red, see text). Here the simulated quenched fractions are computed within the same stellar mass bins as the SAGA observations, although for clarity they are shifted to slightly higher masses. {\it Bottom:} The same as above, but now both simulations and observations have been scaled to a common host K-band magnitude (see eq.~\ref{eq:scale} and associated text).  Accounting for the surface brightness threshold of satellites in SAGA (in particular) and differences in the host K-band magnitude distribution between SAGA and \texttt{ARTEMIS} results in similar quenched fraction trends with stellar mass.}
\label{fig:quenched_frac_SAGA}
\end{figure}

The above comparison does not account for the somewhat different host mass/luminosity distributions of \texttt{ARTEMIS} and SAGA.  Even though the two samples have similar mean K-band luminosities, their distribution of luminosities is not the same (see figure~2 in \citealt{font2021}) and since the abundance of satellites does not scale linearly with K-band luminosity (see below), the stacked quenched fractions will not in general be representative of the mean K-band luminosities of the two samples.  The result will somewhat depend on the detailed distributions.  To account for differences in the host K-band distributions, so that the quenched fractions are representative of the mean K-band luminosities, we first recall the scaling relation between the satellite abundance and the host $K$-band magnitude from \citet{font2021}:

\begin{equation}
N_{\rm sat,~SAGA,\mu~sel.} \propto \biggl(10^{[M_K + 23.5]}\biggr)^{-0.34\pm0.12} \ \ \ .
\label{eq:scale}
\end{equation}

Note that this relation was derived from the \texttt{ARTEMIS} simulations after having applied the SAGA selection criteria described above, including the surface brightness cut, but it also describes the actual SAGA trends very well (see figure~7 of \citealt{font2021}).  

We use eqn.~\ref{eq:scale} to scale the abundance of a given host to the pivot point of $M_K = -23.5$.  As an example, a host with $M_K=-23.0$ has, on average, an abundance that is 0.68 times that of a host with $M_K = -23.5$.  Therefore, to scale the host to the pivot point, we simply multiply by 1/0.68 = 1.48.  For example, if the host had 13 satellites in reality, the scaled abundance would then be $\approx 19$ and we would randomly draw 19 satellites from the list of 13 (some will be repeats), which would be used to compute the scaled quenched fraction.  For a host that is brighter than the pivot scale, the opposite behaviour would result, where we randomly draw a subset of satellites of the host.  In this way, we can statistically scale both the simulations and observations to a common host luminosity.  Note that the quenched fraction of a given host may not change as a result of the above process of random sampling (in fact, it would only change due to noise in the estimate of the quenched fraction), but the contribution of that host to the mean (stacked) quenched fraction will change as a result of its scaled abundance.  Scaling to a common luminosity therefore means giving equal weighting\footnote{In this way, the choice of the actual common host luminosity to scale to is inconsequential, what matters is the equal weighting that results from scaling out the abundance--luminosity relation.} in the quenched fraction to each of the hosts, so that the stacked (mean) quenched fraction is representative of the mean host mass/luminosity\footnote{Another method to achieve consistency between the simulated and observed host distributions is to draw a distribution of hosts from the simulations that match the observed host distribution.  However, such an approach generally requires the simulated sample to be considerably larger than the observed sample, so that the match is accurate.  This is not achievable at a high level of precision with the current \texttt{ARTEMIS} sample, which is of order the same size as SAGA.}.

In the bottom panel of Fig.~\ref{fig:quenched_frac_SAGA} we compare the scaled \texttt{ARTEMIS} and SAGA satellite quenched fractions.  Here we show the 2-sigma uncertainties derived from a binomial beta distribution for both \texttt{ARTEMIS} (black error bars) and SAGA (orange error bars).  We also show with red error bars the maximal systematic error in the SAGA data points using the spectroscopic incompleteness uncertainties from \citet{mao2021}, where we have assumed that the magnitude of these uncertainties is unaffected by the host scaling process.  The agreement in the observed and simulated quenched fractions is similar to that in the top panel, modulo small shifts in the quenched fractions in individual bins and taking into consideration that we are using 2-sigma confidence intervals in the bottom panel for SAGA.  The residual tension that remains could plausibly be explained as a result of using a simple, roughly-determined surface brightness cut (e.g., the appropriate cut may vary from host to host, since the imaging is not uniform) and/or that there are deficiencies in the simulations that make environmental quenching overly efficient.  Regardless, our analysis strongly suggests that selection effects are an important piece of the puzzle.

We note that \citet{karunakaran2021} found a significant tension between the predictions of the Auriga and \texttt{APOSTLE} simulations and the SAGA quenched fraction trend plotted in Fig.~\ref{fig:quenched_frac_SAGA}, but this comparison did not take into account the apparent surface brightness limit of the SAGA data nor differences in the host mass distributions of the observed and simulated systems.  Moreover, \citet{karunakaran2021} also found a tension between the predicted and observed abundances of quenched galaxies.  In \citet{font2021}, we showed that taking into account the SAGA surface brightness limit and differences in the host mass distribution also results in excellent agreement between the predicted and observed total satellite abundances, implying the abundance of quenched galaxies should also be reasonably well recovered.

\begin{figure}
    \centering
    \includegraphics[width=\columnwidth]{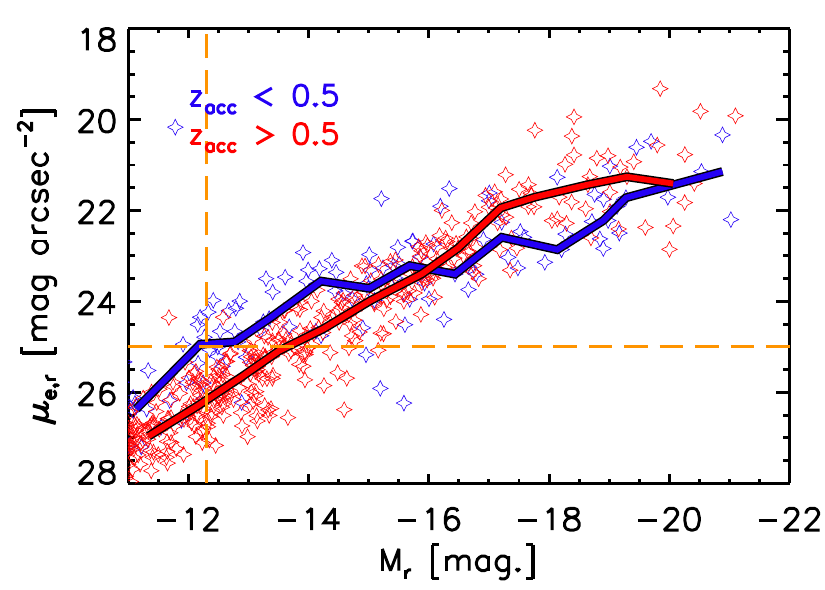}\\
    \includegraphics[width=\columnwidth]{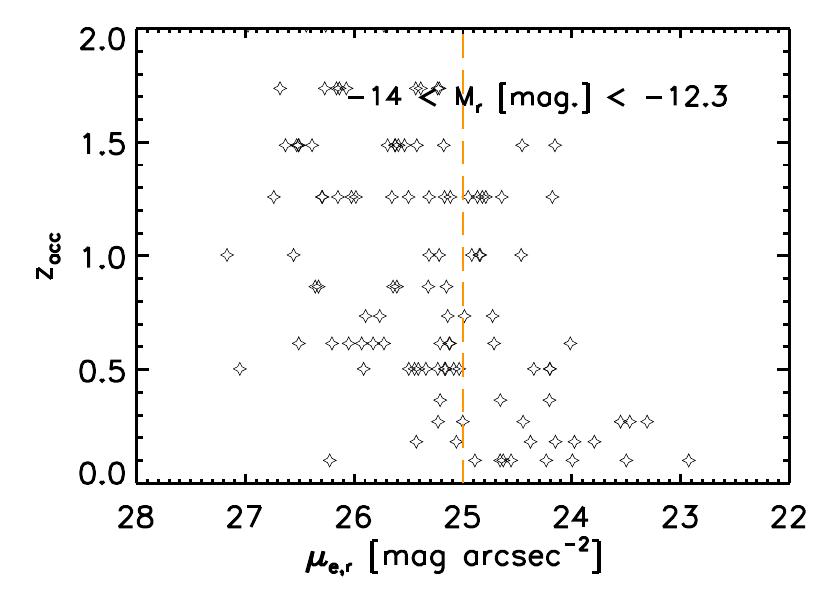}\\
    \includegraphics[width=\columnwidth]{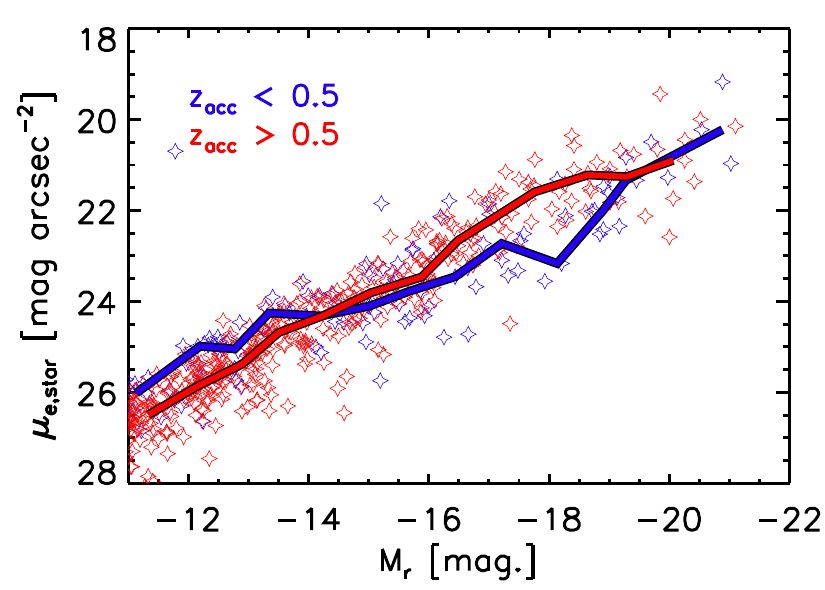}
    \caption{{\it Top:} Magnitude--surface brightness relation of satellites in the simulations, colour-coded according by redshift of accretion, $z_{\rm acc}$.  The solid curves represent the median trends for two cases.  For bright satellites ($M_r < -16$) there is no significant dependence of the relation on the redshift of accretion.  However, for fainter satellites (i.e., near the SAGA selection threshold), recently accreted satellites tend to have increased surface brightness at a given magnitude.  {\it Middle:} The relation between $z_{\rm acc}$ and surface brightness in the magnitude range $-14$ to $-12.3$.  Satellites with surface brightness brighter than about 25 mag arcsec$^{-2}$ are more likely to have been accreted recently.  {\it Bottom:} Magnitude--stellar mass `surface brightness' relation colour-coded according to $z_{\rm acc}$.  Here the surface brightness is estimated using the stellar mass instead of $r$-band luminosity (see text).  There is no significant difference in the magnitude--stellar mass surface brightness relations for satellites of different accretion redshifts, suggesting that dynamical heating does not drive the trend in the top panel.}
    \label{fig:mu_zacc}
\end{figure}

\subsubsection{Physical origin of star-forming selection bias}

We have shown that the detectability, in terms of surface brightness, of relatively low-mass satellites ($M_{\rm star} \la 10^8$ M$_\odot$) can be strongly affected by whether or not the satellite is, or was recently, forming stars.  But what dictates whether a satellite has been forming stars recently and how does this affect the surface brightness?

In the top panel of Fig.~\ref{fig:mu_zacc} we show again the magnitude--surface brightness relations of satellites in the \texttt{ARTEMIS} simulations, but now colour-coded according to the redshift of accretion, $z_{\rm acc}$.  As a reminder, $z_{\rm acc}$ corresponds to the earliest redshift that the satellite first joined the friends-of-friends group of the main progenitor of the host galaxy, using the constructed merger trees.  Here we see that near the selection threshold for SAGA, satellites which were accreted recently ($z_{\rm acc} < 0.5$) tend to have relatively high surface brightnesses and tend to be star-forming recently, whereas satellites that were accreted further in the past tend to be of lower surface brightness and less likely to be forming stars recently.

Selecting satellites in the magnitude range $-14 < M_r < -12.3$, we examine the correlation between surface brightness and the redshift of accretion in the middle panel of Fig.~\ref{fig:mu_zacc}.  This plot shows that satellites with high surface brightnesses ($\mu_{e,r} \ga 25$ mags. arcsec$^{-2}$) tend to have been accreted recently ($z_{\rm acc} \la 0.5$).

One possible interpretation of the results in the top and middle panels of Fig.~\ref{fig:mu_zacc} is that tidal/dynamical heating post accretion lowers the surface brightness by `puffing up' (radially extending) the satellites (see, e.g., \citealt{penarrubia2008}).  This process would be more effective for satellites that have been orbiting the host for longer.  An alternative interpretation is that the quenching of star formation itself strongly reduces the surface brightness and hence detectability.  The quenching of star formation, which is typically centrally concentrated, could result from the removal of cold gas as a result of the ram pressure stripping that occurs as the satellite orbits through the hot gaseous halo of the host.  The removal of cold gas halts the central star formation and surface brightness fading results as the stellar population ages (e.g., \citealt{fang2013}).

To help distinguish between these scenarios, we show in the bottom panel of Fig.~\ref{fig:mu_zacc} the stellar mass `surface brightness', in analogy to the top panel.  Here we have substituted in the star particle stellar mass (in solar masses) for the $r$-band luminosity (in solar $r$-band luminosity) when computing the `surface brightness'.  Essentially this is just the (log) stellar mass surface density but shown in observational units for ease of comparison with the top panel.  The bottom panel does show some dependence of the stellar mass surface brightness on the redshift of accretion at faint luminosities, but it is relatively weak in comparison to the difference in the top panel.  Physically, this implies that the stellar mass distribution has not been strongly altered as a result of accretion, arguing against a dynamical heating scenario as the primary explanation for the trend in the top panel.  Instead, the trends are more readily explained by a surface brightness fading effect that results following ram pressure stripping-induced quenching, perhaps in unison with some degree of stellar mass redistribution.

\subsection{Local Volume (LV)}
\label{sec:LV}

The analysis above suggests that the relatively low quenched fractions inferred for SAGA satellites is at least partially a consequence of observational selection effects, particularly at low stellar masses of $\la 10^8$ M$_\odot$.  By contrast, observations of LG satellites, which typically probe to much fainter surface brightnesses, show higher quenched fractions.  To test that selection effects do indeed account for a large fraction of the apparent discrepancy between SAGA and the LG (rather than appealing to the LG being a very atypical system), we turn now to the LV survey.  While the LV survey contains fewer MW analogues than SAGA, the data is considerably deeper, typically probing down to mean effective surface brightnesses of $\approx 27.6$ mag~arcsec$^{-2}$, thus allowing us to test the above hypothesis. 

For the LV comparison, we use the sample of MW analogues and their dwarf satellites from \citet{carlsten2021a}. From the 9 massive primaries presented in their study, we exclude the two low-mass systems (of roughly LMC-mass), retaining the following systems: NGC 1023 ($M_V=-20.9$), NGC 2903 ($M_V=-20.47$), NGC 4258 ($M_V-20.94$), NGC 4631 ($M_V=-20.24$), M104 ($M_V=-22.02)$, NGC4565 ($M_V=-21.8$) and M51 ($M_V=-21.38 $). 
The LV observations are typically complete for dwarf galaxies with $M_V \simeq -9$, central surface brightnesses of $\mu_{0,V} \simeq 26.5$ mag arcsec$^{-2}$ and projected distances of $\la 150$~kpc \citep{carlsten2021a}.  The central surface brightness limit is converted into a mean effective surface brightness limit of $\mu_{e,V} \simeq 27.6$ mag arcsec$^{-2}$ assuming an exponential light distribution (see \citealt{graham2005}), which is the limit we adopt for the simulations\footnote{In \citet{font2021} we applied a conversion factor of 1.822 to convert the central surface brightness limit of LV to an effective surface brightness limit.  However, as pointed out by the referee of the present study and as described in \citet{graham2005}, a factor of 1.822 converts to the surface brightness \textit{at} the effective radius (assuming an exponential profile), rather than the mean surface brightness within the effective radius, which is the definition we use for \texttt{ARTEMIS} and is also used for SAGA.  The correct factor for converting the central surface brightness to the mean surface brightness within the effective radius is 1.123, yielding an LV limit $\mu_{e,V} \simeq 27.6$ mag arcsec$^{-2}$.  Note that applying the corrected factor to the LV data does not change the conclusion that the SAGA sample appears to have a surface brightness limit of $\mu_{e,r} \simeq 25$ mag arcsec$^{-2}$ and actually slightly improves the agreement of \texttt{ARTEMIS} with LV with regards to satellite abundances in \citet{font2021}.}. In a similar fashion to our comparison with SAGA, we apply an `LV selection' to the simulated dwarf galaxies using the above completeness limits.

\begin{figure}
\includegraphics[width=\columnwidth]{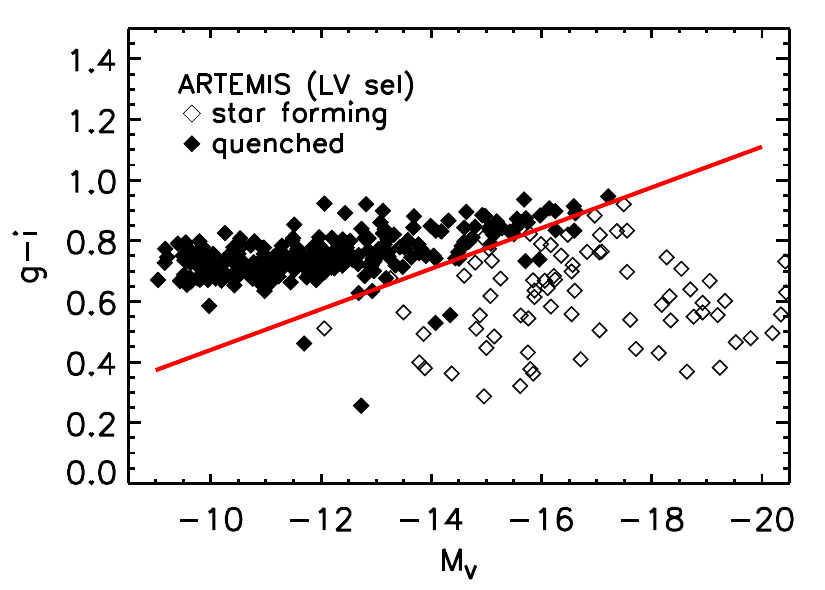}\\
\includegraphics[width=\columnwidth]{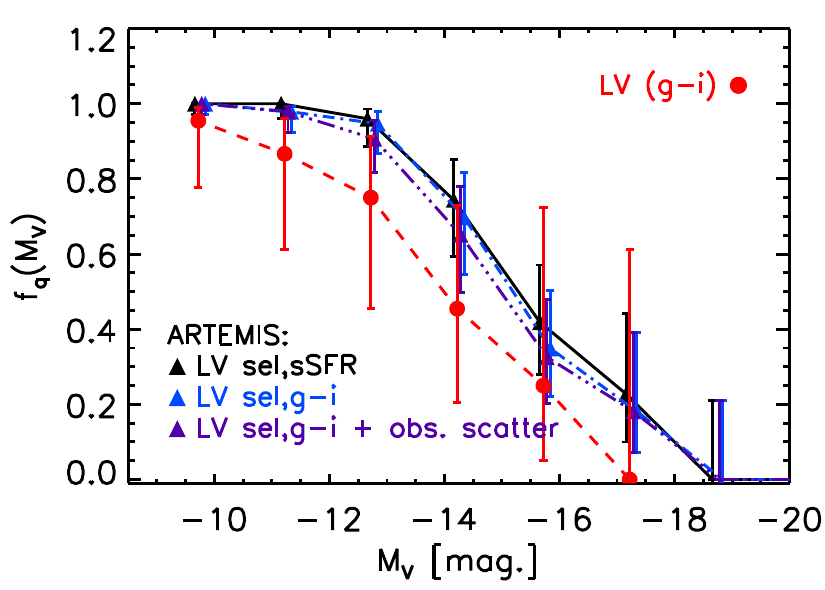}
\caption{{\it Top:} The colour--magnitude relation ($g-i$ versus $M_V$) for simulated dwarf satellites using the `LV selection' (see text). Filled symbols represent the quenched satellites (sSFR$_{\rm inst}=0$). The red dashed line shows the fit of \citet{carlsten2021b} for splitting the observed satellite galaxies by morphological type: early-type galaxies are above this line and late-type below.  This fit separates also separates well the quenched from the star-forming satellites in our simulations. {\it Bottom:} The mean quenched fractions vs.~magnitude in the simulations using the LV selection, designating quenched galaxies using the default criterion ${\rm sSFR}_{\rm inst} =0$ (black) or the colour cut in eq.~\ref{eq:g_i} applied to the raw simulation colours (blue) or to those convolved with observational noise (purple, see text).  For comparison, the mean LV quenched fraction trend is shown which uses eq.~\ref{eq:g_i} to assign quenched status.  The error bars on all data points correspond to the 2-sigma uncertainties derived assuming a binomial beta distribution.  The simulation trends agree well with each other, indicating that colour is an excellent tracer of star formation and both agree reasonably well with the trend inferred from the observed LV satellites, indicating most low-mass satellites are quenched around Milky Way analogues.}
\label{fig:quenched_frac_LV}
\end{figure}

In the following, we will use colours rather than star formation rates to estimate the quenched fractions in the LV samples (both simulated and observed), as measurements of star formation rates in satellite dwarf galaxies in the LV are currently lacking. A more detailed description of how the colours are computed for the simulations is provided in Appendix~\ref{sec:appendixA}.  Generally speaking, we find a strong correlation between galaxy colours and star formation rates in the simulated dwarf sample. Therefore, we expect that the results will not change significantly once SFR data become available.
 
The top panel in Fig.~\ref{fig:quenched_frac_LV} shows the colour--magnitude relation, $g-i$ versus $M_V$, for the simulated dwarf satellites using the LV selection. Filled symbols represent satellites that are not currently forming stars (${\rm sSFR}_{\rm inst} =0$) and are therefore quenched. A clear separation between star-forming and quenched satellites in the simulations is evident in this plot, as expected. The two populations could be easily delineated by a simple linear fit to the simulations.  However, we use instead the observational fit derived by \citet{carlsten2021b}, who used an extended set of observations of dwarf satellites of MW analogues in the LV.  We choose this fit for consistency with the LV observations, although we note that the simulated fit separating the two populations is very similar and adopting it would not change the following results. The observational fit has the form:

\begin{equation}
g-i= -0.067 \times M_V -0.23
\label{eq:g_i}
\end{equation}

\noindent and is shown with a red line in the top panel of Fig.~\ref{fig:quenched_frac_LV}.  This fit does an excellent job of separating the the colours of simulated dwarfs, even though it was derived from the separation of the colours of observed dwarfs {\it based on their morphological type}, i.e., early vs. late-type\footnote{For the LV dataset of \citet{carlsten2021b}, the early-type dwarf galaxies lie above the line and the late-type below it (see figure 18 of that study).}, rather than star formation rates. This suggests the existence of a strong correlation between the star-forming properties and morphologies of dwarf galaxies. This correlation is also discussed in \citet{carlsten2021b} in the context of the observational LV sample, although the observed star formation rates are currently tentative. Further comparisons between the morphologies of dwarf galaxies in the simulations vs.~observations will shed more light on this correspondence, but this is beyond the scope of this study. We note, however, that the simulations have not been fine-tuned in any way to match the observed properties of dwarf galaxies (colours, star formation rates or morphologies), so the close correspondence between the quenched simulated dwarfs and red, early type observed dwarfs (and vice versa, star-forming simulated dwarfs and blue, late-type observed ones) is important, and it further supports the colour bi-modality as a useful tool for designating quenched satellites.
 
\begin{figure}
\includegraphics[width=\columnwidth]{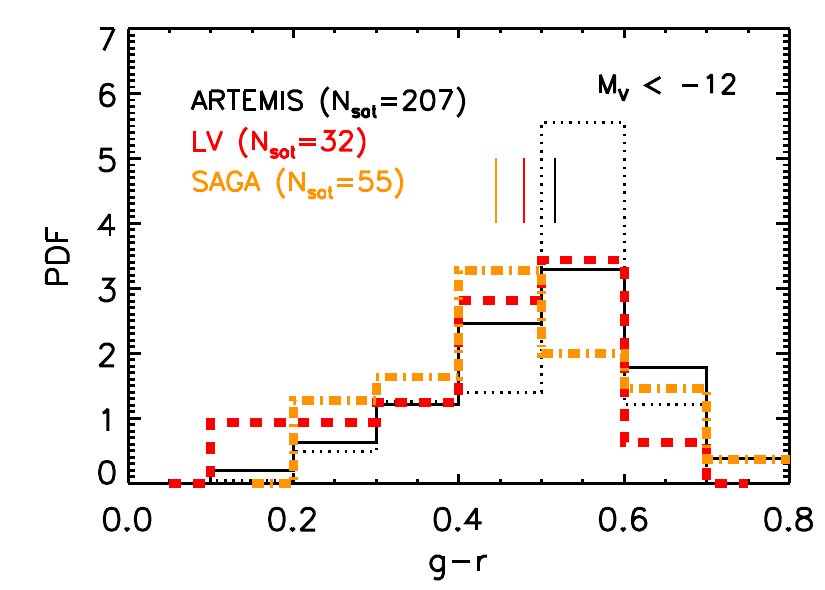}\\
\includegraphics[width=\columnwidth]{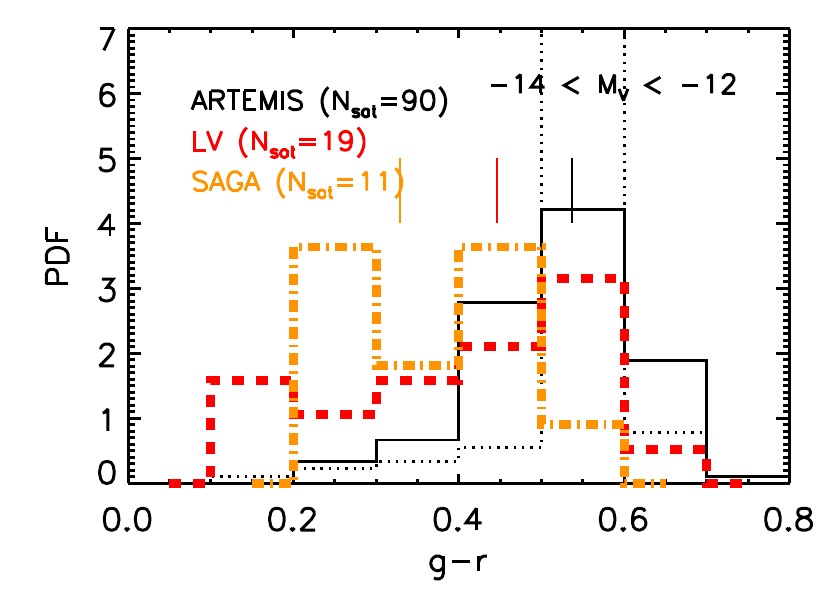}
\caption{Comparison of the distribution of colours ($g-r$) of satellites with $M_V <-12$ (\textit{Top} panel) and with $-14 < M_V < -12$ (\textit{Bottom} panel).  For the simulations and two observational data sets, we consistently select only satellites with projected distances of $<150$~kpc.  The SAGA distribution is generally shifted towards the blue compared with the other two samples, particularly over the magnitude range $-14 < M_V < -12$.}
\label{fig:colours_sats_LV}
\end{figure}

The bottom panel of Fig.~\ref{fig:quenched_frac_LV} shows the mean quenched fraction of satellites in the simulations using the above `LV selection'.  The results using the standard definition for quenched galaxies, sSFR$_{\rm inst}=0$, are shown with black (curve and symbols) and compared with those using the $g-i$ threshold described (shown in blue). 
The two trends predicted from the simulations agree very well, confirming that the colour cut is an appropriate criterion for designating quenched satellites.  We also present a third case for the simulations (in purple), which shows the effect of adding observational noise to the simulated colours at a level of 0.07 dex (assuming a Gaussian distribution), which is approximately the median statistical error on the colour estimates in \citet{carlsten2021b}.  Convolving the raw $g-i$ simulated colours with this noise has the effect of fanning out the sharp red sequence in the top panel, resulting in some satellites at intermediate luminosities being assigned active status rather than their true quenched status.  Consequently, there is a small shift down in the quenched fractions at these luminosities when noise is included, but it does not change the overall picture.

The quenched fractions of the observed satellites in the LV data, averaged over all $7$ MW analogues from \citet{carlsten2021a}, are shown with the dashed red curve in the same panel, where quenched satellites were selecting using the colour cut. Note that here we include both `confimed' and `possible' dwarf satellites from the LV sample, but the results do not change significantly if only `confirmed' satellites are selected. 

Overall, fairly good agreement is found between the observed quenched fractions in the LV survey and those predicted from the simulations, with both pointing to low-mass satellites ($M_V \ga -14$) being highly quenched.  Note that in this comparison we have not accounted differences in the host mass distributions of \texttt{ARTEMIS} and LV, but we highlight that the mean and distribution of host K-band magnitudes are very similar between the two (see \citealt{font2021}) and we have verified that scaling yields similar results and unchanged conclusions.

The quenched fractions inferred from the LV are similar to those obtained for the LG.  They are also in relatively good agreement with the fractions found in SAGA when the effects of the surface brightness limit in the latter are taken into account, along with differences in host mass distributions.
 
As a final check of the impact of the surface brightness limit around $M_r \approx -12$, in Fig.~\ref{fig:colours_sats_LV} we show the distributions of $g-r$ colours for relatively bright satellite galaxies ($M_V<-12$) in the simulations and in the LV and SAGA observations.  The top panel corresponds to all satellites with $M_V < -12$, whereas the bottom panel focuses on the range $-14 < M_V < -12$.
These relatively bright magnitude cuts allow us to directly compare SAGA and LV.  Furthermore, we consistently apply a projected spatial cut of $<150$~kpc.  The colours shown in this figure are in the SDSS bands. As the original colours for the LV observations are in CFHT/MegaCam filters, we have converted these into SDSS bands using the relation $(g - r)^{\rm SDSS} = 1.06(g - r)^{\rm CFHT}$ from \citet{carlsten2021a}.  For some systems in the LV sample where only $g-i$ colours were reported, we use the fit from \citet{carlsten2021a} (see their figure 12) to first transform them into $g-r$.  For \texttt{ARTEMIS} we show two histograms, corresponding to the colour distribution when the raw colours are used (dotted black) and that when then the colours are convolved with an observational scatter of 0.07 dex (solid black).  The coloured vertical solid lines correspond to the median $g-r$ values of the three samples (LV, SAGA, and \texttt{ARTEMIS} with obs.~scatter).

For satellites with $M_V < -12$ (top panel), the simulated and observed LV distributions are in fairly good agreement (particularly when the effect of observational scatter is included), both with a $g-r$ peak around $0.55$.  The satellites in the SAGA hosts are also similar, but typically have somewhat bluer colours, with a peak around $0.45$. Narrowing the comparison to the range $-14 < M_V < -12$, this difference is increased, with the SAGA satellites tending to be significantly bluer than those in LV and especially with respect to \texttt{ARTEMIS}, although we caution that the number of satellites in the comparison is significantly reduced for this narrower magnitude range.  The fact that there is a difference in the colours between SAGA and LV even after the samples have been brought into consistency (i.e., same $M_V$ limit and spatial coverage), suggests that the additional surface brightness limit in SAGA, discussed in Section~\ref{sec:SAGA}, is likely responsible for this difference.  Consistent with this picture is that, over the  magnitude range $-14 < M_V < -12$, there are actually more satellites in LV (19) than in SAGA (11), in spite of there being approximately 5 times as many hosts in SAGA relative to LV in this comparison.  As demonstrated by \citet{font2021}, this difference in the abundance of satellites can also be neatly accounted for when incorporating a plausible surface brightness limit.

\section{Discussion and Conclusions}
\label{sec:concl}

Using the \texttt{ARTEMIS} hydrodynamical simulations, we have examined the quenching of satellites around Milky Way-mass hosts and made careful, consistent comparisons with the observed quenched fractions of satellites in the Local Group and those around Milky Way analogues in the SAGA and Local Volume surveys.  The conclusions of this study can be summarised as follows:

\begin{itemize}

    \item The simulations predict that the dwarf satellites of MW analogues are significantly quenched, particularly at low masses.  For example, on average, $>60\%$ of dwarfs with ${\rm M}_{*} < 10^8 \, {\rm M}_{\odot}$ are predicted to be quenched.  This agrees well with that observed for the combined Local Group system (Fig.~\ref{fig:quenched_frac_LG}).  However, there is also significant system-to-system scatter, in agreement with differences seen in the MW and M31 observations. 
 
    \item The quenched fractions derived from relatively shallow observations, compared to what can be obtained for the Local Group, can be potentially biased due to selection effects.  Specifically, we have shown that satellites that are (or recently were) star forming tend to have higher surface brightnesses at fixed magnitude in the range $M_V \approx -12$ to $-14$ (see Fig.~\ref{fig:mu_SFR_mag}), implying that star-forming satellites are more easily detected in observations in this magnitude range.
    
    \item The above magnitude range is particularly relevant for SAGA, as it is near the completeness limit of that survey and represents the range where the majority of satellites live.  Comparison to the deeper LV survey suggests a surface brightness limit of $\sim$ 25 mag~arcsec$^{-2}$ in SAGA.  Applying such a surface brightness limit to the simulations results in nearly a factor of two drop in the derived quenched fractions at low masses (see top panel of Fig.~\ref{fig:quenched_frac_var}), as the excluded low surface brightness satellites are preferentially quenched.
   
    \item The derived mean quenched fractions also depend on the properties of host galaxies, primarily on the total host mass. Namely, at fixed dwarf stellar mass, higher mass systems contain higher fraction of quenched dwarfs (see the bottom panel of Fig.~\ref{fig:quenched_frac_var}). Taking into account this dependency is required in order to make consistent comparisons between observations and simulations of MW analogues, as these samples include a range of host masses.

    After applying SAGA's specific selection criteria to the simulated sample, including the additional surface brightness limit, and correcting for the differences in the simulated and observed host mass distributions, we find that the theoretical predictions for the quenched fractions agree reasonably well with (though slightly above) those inferred from observations (Fig.~\ref{fig:quenched_frac_SAGA}), in contrast to the findings of \citet{karunakaran2021} who found a significant tension between SAGA and the predictions of simulations. The differing theoretical results derived by \citet{karunakaran2021} using the Auriga and \texttt{APOSTLE} simulations compared to our \texttt{ARTEMIS} predictions is primarily due to differences in the applied satellite selection criteria, rather than large intrinsic differences in the simulation predictions.  Specifically, we have included a plausible surface brightness limit and accounted for different host mass distributions when comparing to SAGA, whereas these effects were not included in the analysis of Auriga and \texttt{APOSTLE}. 
    
    \item In \texttt{ARTEMIS} we have shown that satellites in the magnitude range $M_V \approx -12$ to $-14$ with relatively high surface brightness ($\mu_{e,r} \ga 25$ mag arcsec$^{-2}$) tend to have been accreted recently, whereas those with relatively low surface brightness were accreted earlier (see Fig.~\ref{fig:mu_zacc}).  The lack of a strong trend in stellar mass surface density suggests that it is not tidal heating/stripping that drives the scatter in surface brightness (though it may contribute at some level), but rather the quenching itself.  We suggest that disc fading associated with ram pressure stripping-inducing quenching is the most likely driver. 
    
    \item We made additional predictions for the population of quenched satellites of MW-analogues in the LV survey, which has surface brightness limitations between those in SAGA and in the LG. The quenched fractions of faint ($M_V \la -12$) satellites in the LV sample are expected to be high, with similar values as found in, and predicted for, the LG.  Using the $g-i$ colour as an indicator of star formation activity, we have compared the predictions of \texttt{ARTEMIS} with the LV observations, finding relatively good agreement (Fig.~\ref{fig:quenched_frac_LV}), indicating that low-mass satellites are indeed likely to be highly quenched around MW-mass hosts.
    
    \item We have shown that over the magnitude range $M_V \approx -12$ to $-14$, the colour of SAGA satellites appears to be significantly bluer than that of LV satellites (and \texttt{ARTEMIS} satellites; see Fig.~\ref{fig:colours_sats_LV}).  This trend, combined with the substantial difference in satellite abundance of the same magnitude range, suggests that relatively faint, red satellites may be preferentially selected against in SAGA.

\end{itemize}

Our findings therefore illustrate that there is no significant tension in the predicted quenched fractions from $\Lambda$CDM-based simulations and that observed for Milky Way-mass hosts in the local Universe.  It further underscores the importance of making like-with-like comparisons between theory and observations which take into account the relevant observational selection criteria.  

\section*{Acknowledgments}
We thank the referee for the constructive comments
that have significantly improved our paper.  We also thank Ananthan Karunakaran and Kristine Spekkens for helpful feedback.  This project has received funding from the European Research Council (ERC) under the European Union's Horizon 2020 research and innovation programme (grant agreement No 769130).  STB and SGS acknowledge STFC doctoral studentships. This work used the DiRAC@Durham facility managed by the Institute for Computational Cosmology on behalf of the STFC DiRAC HPC Facility. The equipment was funded by BEIS capital funding via STFC capital grants ST/P002293/1, ST/R002371/1 and ST/S002502/1, Durham University and STFC operations grant ST/R000832/1. DiRAC is part of the National e-Infrastructure.

\section*{Data availability}
The data underlying this article may be shared on reasonable request to the corresponding author.




\bibliographystyle{mnras}
\bibliography{references} 


\appendix

\section{Star forming status criterion}
\label{sec:appendixB}

\begin{figure}
\includegraphics[width=\columnwidth]{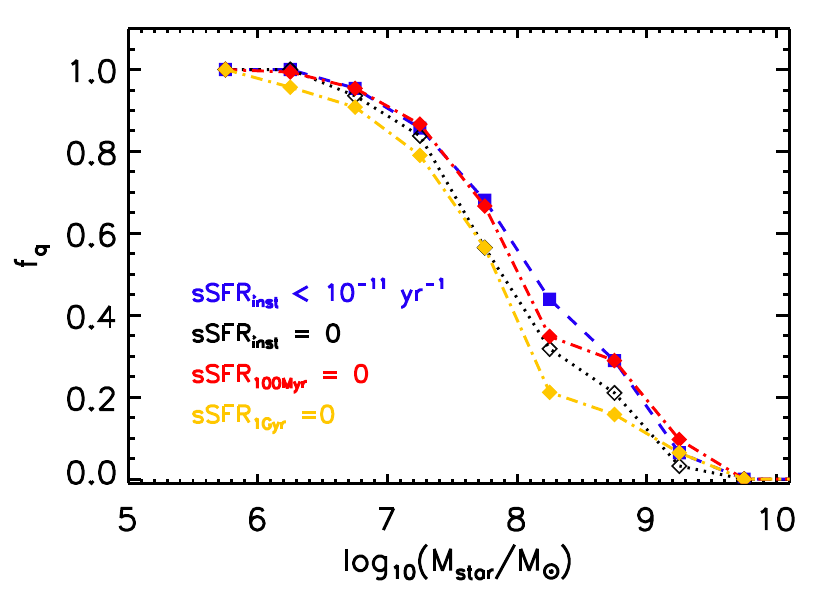}\\
\caption{Quenched fractions in the simulations, using different criteria for the definition of `quenched': those satellites that have instantaneous sSFR$_{\rm inst}< 10^{-11}\, {\rm yr}^{-1}$ at present time (blue dashed curve) or $0$ (black dotted curve, the default criterion); or those satellites that have not formed stars over the past $100$~Myr or $1$~Gyr (red and gold dot-dashed curves, respectively).}
\label{fig:quenched_frac_sfr}
\end{figure}

Here we explore the impact of changing the criterion for assigning star forming status.  Fig.~\ref{fig:quenched_frac_sfr} shows the quenched fractions of simulated satellites using different definitions for what constitutes a quenched galaxy. Specifically, we compare the default case (quenched if currently not forming any stars, sSFR$_{\rm inst} =0$) with the following definitions:

\begin{itemize}
\item quenched if sSFR$_{\rm inst}> 10^{-11}\, {\rm yr}^{-1}$
\item quenched if sSFR$_{100\, \rm{Myr}}=0$
\item quenched if sSFR$_{1\, \rm{Gyr}}=0$.
\end{itemize}

The first case corresponds to a commonly adopted criterion in extragalactic environmental studies of relatively massive satellites.  The last two cases use the time-averaged star formation rate either over the past 100 Myr or past 1 Gyr, as described in Section \ref{sec:SAGA}. 

The change in sSFR$_{\rm inst}$ threshold, from 0 (black dotted curve) to $10^{-11}$ (blue dashed) results in a small increase in the quenched fractions, particularly at the high-mass end where a few dwarf galaxies are star-forming at low specific rates. Changing the sSFR criterion from present-day (sSFR$_{\rm inst}$) to one averaged over a limited time (sSFR$_{\Delta t}$) also has a modest effect. For example, using the time-averaged sSFR over the past 1 Gyr (orange curve) results in a slightly decreased quenched fractions across all dwarf mass range. The case of $\Delta t=100$~Myr lies between these two cases.  

Note that, in terms of comparison with the observations, the instantaneous rate is likely the most consistent choice for comparison with $H_{\alpha}$, whereas the sSFR averaged of a timescale of 100 Myr may be more appropriate for a comparison with UV-based estimates of the SFR.  

\section{Computing colours of simulated satellites}
\label{sec:appendixA}

\citet{font2020} showed that \texttt{ARTEMIS} predicts total metallicities ($Z$) that are too high compared to observed low-mass galaxies (see figure~2 of that study), implying that the predicted colours for low-mass galaxies would be redder than observed, a result that we have confirmed.  A closer examination of the simulations reveals that, while the total metallicities are high compared to observations, the [Fe/H] abundance agrees relatively well with current data.  As Oxygen (O) dominates the total metal abundance, this could suggest that either the adopted O yields are too high, or else the feedback requires a preferential O mass-loading in order to prevent it from ending up in stars.  In any case, since the [Fe/H] abundances reproduce the observations relatively well, we use it as our total metallicity indicator.  Specifically, we use [Fe/H] from the simulations and then assume solar abundance ratios (an approach commonly used in observational analyses) to derive an estimate of the total metallicity.  The metallicity, age, and initial mass are used with the above stellar population models to derive luminosities in a variety of bands via the PARSEC SSP models (see Section \ref{sec:sims}).  Note that we neglect the effects of dust attenuation when estimating luminosities and colours.

We note that our comparison to SAGA is not affected by uncertainties in metallicities, colours or stellar populations, since we use the SFR as an indicator for quenching of galaxies.  The colours are mainly used for cross-checking our SFR results and are only relevant for the LV comparison (see Section \ref{sec:LV}).  Furthermore, they are primarily used to separate the bi-modal populations, implying that only the shape of the colour distribution (and not the absolute colours themselves) is of consequence.

\bsp	
\label{lastpage}
\end{document}